\documentclass[final,5p,times,twocolumn]{elsarticle}

\usepackage{graphicx}

\journal{Nuclear Instruments Method}

\begin{document}
\begin{frontmatter}



\title{Development of portable NMR polarimeter system for polarized HD target}


\author{T.~Ohta$^{a,*}$, M.~Fujiwara$^{a}$, K.~Fukuda$^{b}$, H.~Kohri$^{a}$, T.~Kunimatsu$^{a}$, C.~Morisaki$^{a}$, S.~Ono$^{a}$, M.~Tanaka$^{c}$, 
K.~Ueda$^{a}$, M.~Uraki$^{a}$, M.~Utsuro$^{a}$, S.Y.~Wang$^{d,e}$, and M.~Yosoi$^{a}$}

\address{
$^{a}${\it Research Center for Nuclear Physics, Mihogaoka 10-1, Ibaraki, Osaka 567-0047, Japan} \\
$^{b}${\it Kansai University of Nursing and Health Sciences, Shizuki Awaji 656-2131, Japan}\\
$^{c}${\it Kobe Tokiwa University, Ohtani-cho 2-6-2, Nagata, Kobe 654-0838, Japan}\\ 
$^{d}${\it Institute of Physics, Academia Sinica, Taipei 11529, Taiwan}\\
$^{e}${\it Department of Physics, National Kaohsiung Normal University,
 Kaohsiung 824, Taiwan}\\
}

\begin{abstract}
A portable NMR polarimeter system has been developed to measure the polarization of a polarized Hydrogen-Deuteride (HD) target for hadron photoproduction experiments at SPring-8. 
The polarized HD target is produced at the Research Center for Nuclear Physics (RCNP), Osaka university and is transported to SPring-8. 
The HD polarization should be monitored at both places.
We have constructed the portable NMR polarimeter system by replacing the devices in the conventional system with the software system with PCI eXtensions for Instrumentation (PXI). 
The weight of the NMR system is downsized from 80 kg to 7 kg, and the cost is reduced to 25\%. 
We check the performance of the portable NMR polarimeter system.
The signal-to-noise ($S/N$) ratio of the NMR signal for the portable system is about 50\% of that for the conventional NMR system. 
This performance of the portable NMR system is proved to be compatible with the conventional NMR system for the polarization measurement.
\end{abstract}


\end{frontmatter}


\section{Introduction}
\label{intro}
After the completion of the SPring-8/LEPS facility in 2000, a linearly polarized photon beam at $E_{\gamma}$=1.5-2.4 GeV has been used for studying hadron photoproduction reactions. 
Introducing a polarized target is believed to play a crucial role in understanding the hadron structures~\cite{Titov97} and reaction mechanisms~\cite{Mibe05, Kohri10}, and would result in upgrades of the LEPS experiments. 
For this purpose, we have developed a polarized Hydrogen-Deuteride (HD)~\cite{Fujiwara03} target for future experiments at SPring-8/LEPS. 

After the important work on the relaxation mechanism for polarized protons in solid HD~\cite{Bloom57,Hardy66}, Honig first suggested that the HD molecule can be used as a frozen spin polarized target~\cite{Honig67}.
Thanks to longstanding efforts at Syracuse~\cite{Honig89,Honig95}, BNL~\cite{Wei00,Wei01,Wei04}, and ORSAY~\cite{Breuer98,Rouille01,Bassan04,Bouchigny05,Bouchigny09}, the HD target has been firstly used for the actual experiment at LEGS~\cite{Holbit09}, and will be used both at JLab~\cite{Sandorfi} and at SPring-8~\cite{Kohri10-1} in near future. 
In order to achieve high polarizations of proton and deuteron in the HD target, we employ the static method ("brute force" method) at low temperature (10 mK) and at high magnetic field (17 T). The expected polarization of proton is 94\% at maximum. 
In order to polarize protons in the frozen HD target, Honig~\cite{Honig67} applied the innovative idea that the HD polarization gradually grows up in the spin-flip process between HD molecules and a small amount of ortho-H$_{2}$ with spin 1 by making use of the mechanism originally clarified in earlier years by Motizuki {\it et al.}~\cite{T. Moriya,K. Motizuki}.
After a long period of cooling, ortho-H$_{2}$ molecules are converted to para-H$_{2}$. Since the para-H$_{2}$ has a total spin 0, the spin-flip process will be ended when all the ortho-H$_{2}$ molecules are converted to the para-H$_{2}$.
If ortho-H$_{2}$ does not remain in the HD sample, the relaxation time of hydrogen polarization becomes very long even at a temperature higher than 1 K after an aging process for 2-3 months.

A polarized HD target is produced at RCNP, and then it is transported to the LEPS facility at SPring-8 which locates at a distance of about 130 km from RCNP.
The HD polarization should be monitored by using the same NMR polarimeter system at RCNP and at SPring-8. The polarization must be determined with a precision of 10\% for distinguishing the theoretical model calculations~\cite{Titov97}. For this purpose, a portable NMR system is desired. 

In the 1960's, NMR measurements for the development of polarized targets
were carried out by using a diode demodulator and a tuned RF amplifier. Poor $S/N$ ratio and linearity of these devices restricted the precision
of the measurements.
At the end of the 1960's, the performance and reliability of the NMR measurements
greatly improved due to the development of the synchronous demodulation technique,
called the phase-sensitive demodulator (PSD). A lock-in amplifier was developed by combining the PSD
with the tuned RF amplifier and widely used for the NMR measurements.
Although the devices for the NMR measurements become downsized,
they are still heavy and large for a long-distance transportation.
We newly developed a portable NMR system by employing cutting-edge digital
technologies for the fast ADC and for software logic circuits with a usual laptop computer.
This paper describes the basic concept and the performance
of the portable NMR system.

\section{Polarization measurement method}
\begin{figure}[htb]
  \begin{center}
    \includegraphics[width=80mm]{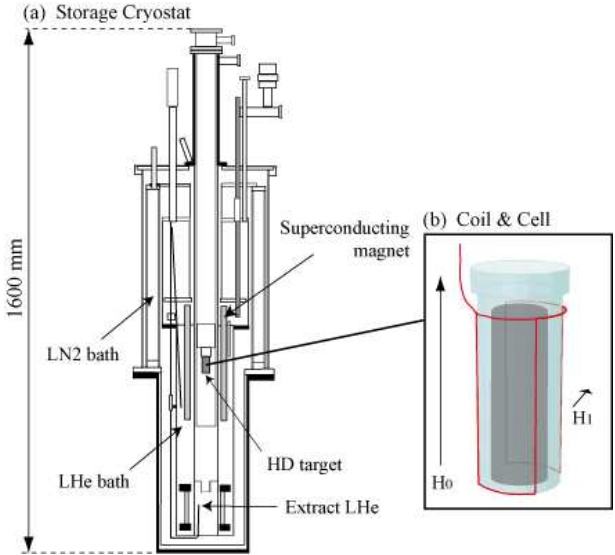}
  \end{center}
  \caption{(a) The structure of the storage cryostat (SC) produced by Oxford Instruments and (b) the details of the HD cell, coil, and its  support frame. The geometry of the HD target cell is 2.5 cm in diameter and 5 cm in length for the LEPS experiments. The coil winding is illustrated with the directions of the magnetic field $H_{0}$ of  the 
superconducting magnet and the applied RF field $H_{1}$. }
  \label{fig:SC_Coil.eps}
\end{figure}

Fig.~\ref{fig:SC_Coil.eps}(a) schematically shows the storage cryostat (SC) which is used for solidifying HD and for keeping the polarization of the HD target during the transportation from RCNP to SPring-8.
The SC mainly consists of a liquid nitrogen (LN$_2$) bath, a liquid helium (LHe) bath, and a superconducting magnet. 
The SC is equipped with a needle valve to control the HD temperature in the range of 1.5-30 K by pumping evaporated He gas from the LHe bath. 
The superconducting magnet cooled by LHe provides the maximum magnetic field of 2.5 T. 
The homogeneity, $\Delta$B/B, of the magnetic field is $1.0\times10^{-4}$  in the central region over 300 mm long and 30 mm diameter. 
The HD gas is solidified in a cell made of Kel-F$\textregistered$ (PCTFE:Poly-Chloro-Tri-Fluoro-Ethylene), a hydrogen-free material. 

The set-up of the coil support frame around the target cell  in the SC is shown in Fig.~\ref{fig:SC_Coil.eps}(b).
We apply a single coil method for NMR measurements~\cite{Fukushima,Klein}. 
A Teflon coated silver wire with a diameter of 0.1 mm is wounded as a saddle coil by 1 turn on the coil support frame which is also made of Kel-F$\textregistered$.

The polarization of a nucleus with a spin of 1/2 is defined as $P=(N_{+} - N_{-})/(N_{+} + N_{-})$, where $N_{+}$ and $N_{-}$ are the numbers of sub-states $m=+1/2$ and $m=-1/2$, respectively.
The absorption strength of the RF power in the NMR coil is proportional to magnetization which is also proportional to the polarization~\cite{Abragam61}. 
Therefore, the proportionality between the polarization and the absorption strength can be used for determining the polarization degree.



\begin{figure}[h]
  \begin{center}
    \includegraphics[width=80mm]{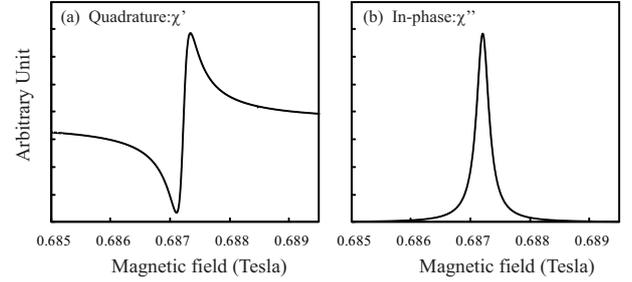}
  \end{center}
  \caption{Expected NMR signals for hydrogen in HD molecules. (a) The quadrature signal corresponds to the dispersion function given in Eq. (\ref{eqn:chi''}). (b) The strength of in-phase signal proportional to the polarization degree which corresponds to the absorption function given in Eq. (\ref{eqn:chi'}).  These NMR signals are simulated under the condition that the frequency is fixed at 29.26 MHz, the magnetic field is swept from 0.67 to 0.75 Tesla, and  $T_{1}$ and $T_{2}$ are assumed to be 1 second.}
  \label{fig:In_Quad.eps}
\end{figure}

The details about the NMR absorption functions are given by Abragam~\cite{Abragam61}. 
The susceptibility $\chi=\chi'-i\chi''$ is defined by the two functional forms  
as
\begin{eqnarray}
\chi'&= &\frac{1}{2}\cdot\frac{\omega_{0}\Delta\omega T_{2}^{2}}{ 1+(T_{2}\Delta\omega)^{2}+\gamma^{2}H_{1}^{2}T_{1}T_{2}} \label{eqn:chi'}\\
\chi''&= &\frac{1}{2}\cdot\frac{\omega_{0}T_{2}^{2}}{ 1+(T_{2}\Delta\omega)^{2}+\gamma^{2}H_{1}^{2}T_{1}T_{2} }, \label{eqn:chi''}
\end{eqnarray}
where $T_{1}$ is the longitudinal relaxation time, $T_{2}$  is the transverse relaxation time,
${\gamma}$ is a gyro-magnetic ratio, ${\omega}_0$ is  resonance frequency, $\omega$ is 
sweep frequency,  $\Delta\omega$ is defined as $\Delta\omega=\omega_0-\omega$, and 
$H_1$ is the rotating field amplitude.
Two responses shown in Fig.~\ref{fig:In_Quad.eps} correspond to the dispersion and  absorption functions  obtained from Eqs. (\ref{eqn:chi'}) and (\ref{eqn:chi''}) which are derived from the Bloch equation~\cite{Bloch46}.  
The polarization of the target is proportional to the strength of the absorption function $\chi''$. Thus polarization P can be expressed as follows~\cite{M. Goldman, G. R. Court, K. Kondo}.
\begin{equation}
P\sim \int ^{\infty }_{0}\chi''(\omega)d\omega.
\end{equation}

\section{Development of  the portable NMR polarimeter system}
\subsection{Hardware in the conventional system}
\begin{figure}[htb]
  \begin{center}
    \includegraphics[width=80mm]{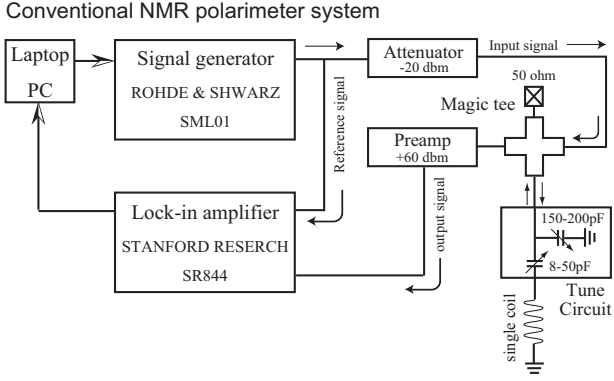}
  \end{center}
  \caption{Schematic drawing of the conventional NMR polarimeter electronics.}
  \label{fig:NMR_diagram.eps}
\end{figure}

The conventional NMR polarimeter system consists of a signal generator, an oscilloscope, 
a lock-in amplifier and a network analyzer. The network analyzer is used for minimizing power reflection of the NMR circuit by tuning variable capacitors at off-resonance frequencies. 
The oscilloscope is used to observe the signals returned from the NMR coil.
The lock-in amplifier is used to pick up small RF signals with a frequency equal to the frequency of the input signal. Even if a noise level is several thousand times higher than a true tiny NMR signal, a signal with a specific frequency can be extracted by using a phase sensitive detection method. Noises with frequencies other than the reference frequency are rejected. As a result, we can greatly reduce the effect of noises in the NMR measurement.
As shown in Fig.~\ref{fig:NMR_diagram.eps}, we introduced a special circuit to cancel output signals in the case of the non-resonance. 
The sinusoidal wave output from the signal generator is divided into two components.
One is sent to the NMR coil and the other is sent to the lock-in amplifier as a reference signal.
When the nuclear magnetic resonance occurs, the HD target absorbs RF energies from the coil, and the absorption signal is measured with the lock-in amplifier when magnetic field (or frequency) is swept.
Essentially, these devices are able to be constructed from two functions to generate RF signals and to digitize the signals. 
If these two functions are integrated into one device, it is expected that the polarimeter system is downsized and its cost is greatly reduced.

\subsection{Hardware development in the portable system}
\begin{figure}[!htb]
  \begin{center}
    \includegraphics[width=80mm]{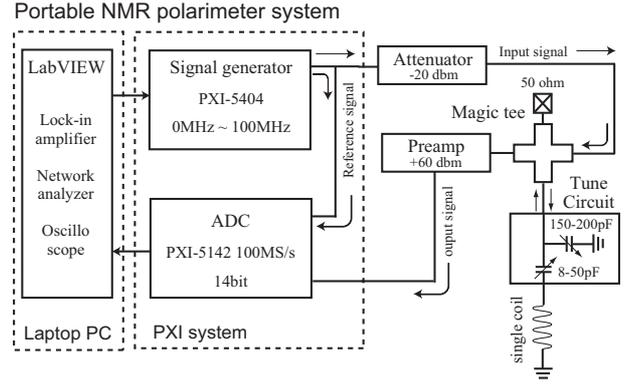}
  \end{center}
  \caption{Schematic drawing of the portable NMR polarimeter system: The signal generation and data acquisition are realized by using PXI-5404 module (signal generator) and PXI-5142 (ADC) module with the software (LabVIEW). }
  \label{fig:NMR_PXI_diagram.eps}
\end{figure}

\begin{figure}[!htb]
  \begin{center}
    \includegraphics[width=80mm]{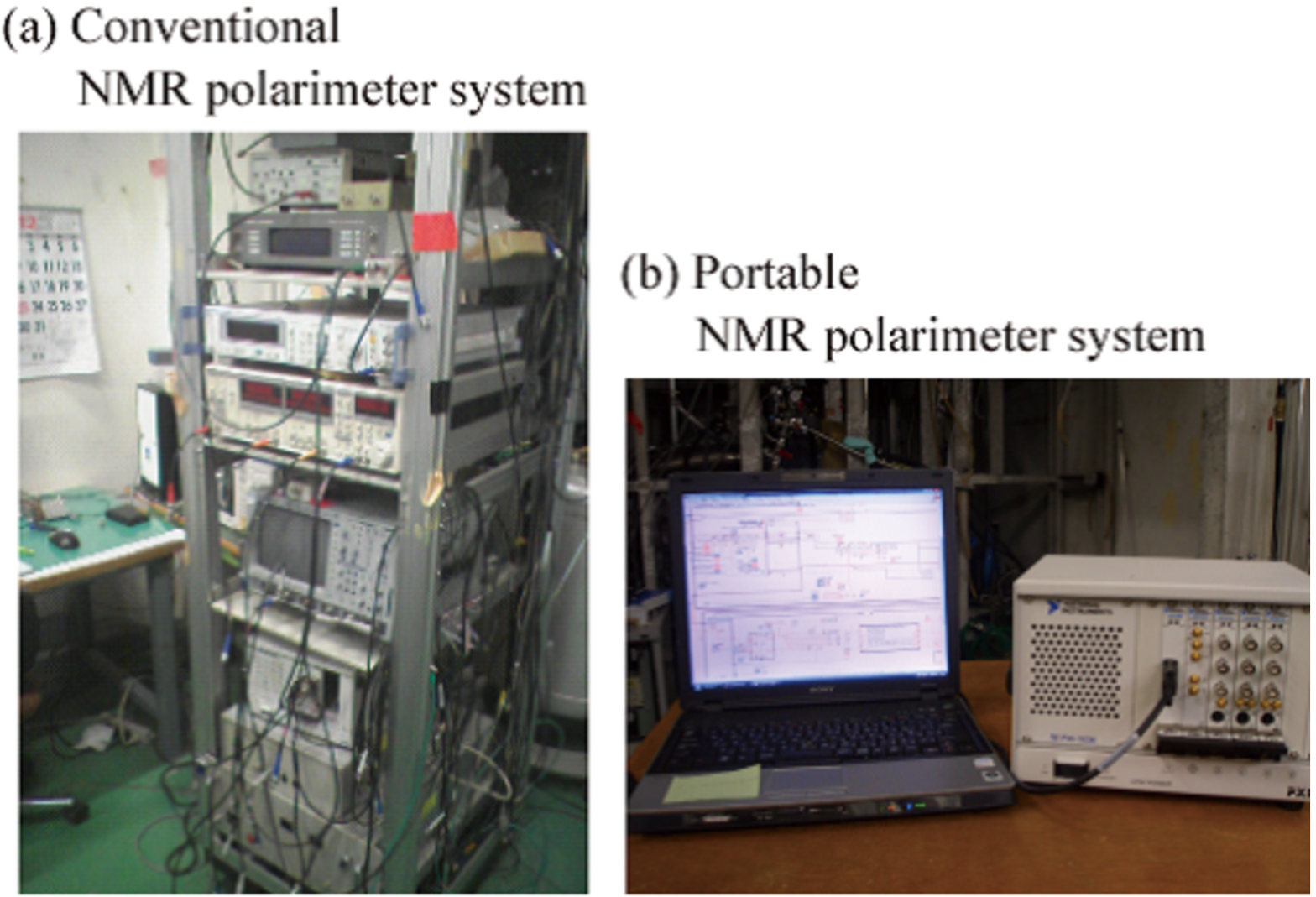}
  \end{center}
  \caption{(a) The conventional NMR polarimeter system installed in a rack. Essentially, we need to prepare a signal generator, an oscilloscope, 
a lock-in amplifier, and a network analyzer. (b) The portable NMR polarimeter system consisting of a laptop PC and a PXI system. }
  \label{fig:Sizedown.eps}
\end{figure}

We have constructed the portable NMR polarimeter by using "PCI eXtensions for Instrumentation (PXI)"~\cite{PXI}.
This system consists of PXI-1036 (chassis), PXI-8360 (connection between PC and PXI), PXI-5404 (signal generator), and PXI-5142 (ADC) which are produced by National Instruments Company.
The hardwares used in the present work are listed in Table~\ref{Tab:Hardwares}.
The aforementioned functions necessary for NMR measurements are virtually implemented and are 
realized using the LabVIEW software. 
The schematic drawing of the portable NMR polarimeter system is shown in Fig.~\ref{fig:NMR_PXI_diagram.eps}.
\begin{table}[h]
\caption{Hardware of the portable NMR polarimeter system}
\label{Tab:Hardwares}
 \begin{center}
  \begin{tabular}{|c|c|c|c|c|}
    \hline
    Product name  & Function   &  Specification    \\
    \hline
    PXI-1036  & Chassie   &  host 1slot, module 5slot    \\
    \hline
    PXI-8360  & PCI bus   & throughput 110MB/s    \\
    \hline
    PXI-5404  & Signal generator   & 16bit, 0-100MHz    \\
    \hline
    PXI-5142  & ADC   & 14bit, 2GS/s   \\
    \hline
  \end{tabular}
 \end{center}
  \end{table}
Fig.~\ref{fig:Sizedown.eps} shows the photographs of the conventional and portable NMR systems.
Table~\ref{Tab:Comparison} compares some specifications of the conventional and portable NMR systems. 
Downsizing the weight of the NMR system from 80 kg to 7 kg and reducing the cost to 25\% were successfully achieved. 
\begin{table}[!htb]
 \caption{Comparison of weight, size, and cost for the conventional and portable NMR systems. The width, depth, and height of the conventional system are 500 mm, 500 mm and 1000 mm, respectively. Those of the portable system are 200 mm, 200 mm and 250 mm, respectively. The total sizes are compared in percentage. The weight of the portable NMR system does not include the laptop PC}
\label{Tab:Comparison}
 \begin{center}
  \begin{tabular}{|c|c|c|c|}
    \hline
    &weight& size & cost\\
    \hline
    Conventional system&80 kg&100\% & \$60,000\\
    \hline
    Portable system&  7.1 kg& 4.0\% &\$15,000\\
     \hline
        \end{tabular}
 \end{center}
 \end{table}

\subsection{Software development in the portable system}
We developed a software LabVIEW program on the laptop PC to separate a specific NMR in-phase component from other frequencies and noises.
In the measurement of the polarization with the NMR method, we need only the in-phase component
since the polarization degree is proportional to the strength of the in-phase signal.
In every measurements, the phase is adjusted to be in-phase.
If the phase is apart from in-phase, we can not obtain an accurate polarization degree, because the
susceptibility has two components ${\chi}'$ and ${\chi}''$.
In case of the actual measurement, in-phase and quadrature components are simultaneously measured
for rebuilding the in-phase signal.
By making a phase rotation of the measured data in a complex plane, we can reproduce a correct
in-phase component in the off-line analysis.
\begin{figure}[hb]
  \begin{center}
    \includegraphics[width=80mm]{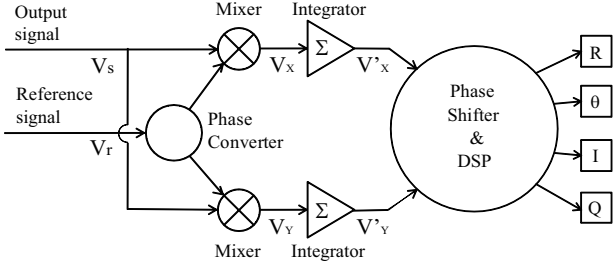}
  \end{center}
  \caption{Schematic drawing of software diagram for the portable NMR polarimeter system. I: NMR in-phase signal, Q: NMR quadrature signal, R: Amplitude of signal, $\theta$: signal phase.}
  \label{fig:Soft_diagram.eps}
\end{figure}
As shown in Fig.~\ref{fig:Soft_diagram.eps}, R, $\theta$, I and Q components are obtained from the output signal $V_{s}=A_{s}\sin(\omega t+\theta_{s})$ and the reference signal $V_{r}=A_{r}\sin(\omega t+\theta_{r})$. 
Here $A_{r}$ and $A_{s}$ are the signal amplitudes, and $\theta_{r}$ and $\theta_{s}$ are the signal phases.

\begin{figure}[h]
  \begin{center}
     \includegraphics[width=60mm]{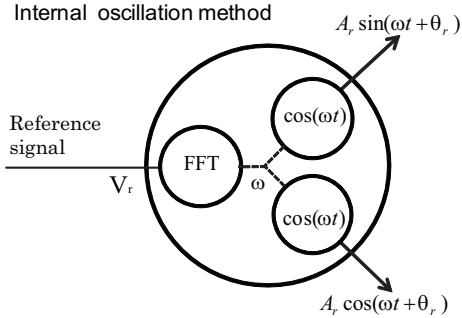}
  \end{center}
  \caption{Method for phase conversion. To produce a signal with quadrature phase from the reference signal, sine and cosine functions with a frequency of the reference signal are generated. 
  The frequency is determined from the optimum peak position in the FFT spectrum for the reference signal.}
\label{fig:Phase_Converter_method.eps}
\end{figure}

To obtain the two signals, $I$ and $Q$, the reference signal is divided into two signals with sine and cosine functions by using a phase converter.
Phase conversion is performed with the software. 
For generating the cosine function from the sine function, we used "internal oscillation method" (see Fig.~\ref{fig:Phase_Converter_method.eps}). 
In this method, the reference signal is firstly converted to a spectrum as a function of frequency by applying FFT (Fast Fourier Transformation) method. 
We determine the optimum frequency from the peak location in the FFT spectrum.
By using the optimum frequency determined, two signals with waveform of $\sin(\omega t +\theta_{r})$ and $\cos(\omega t +\theta_{r})$ are generated. 
Digital data for these two waveforms are stored in a memory over the period of the Fourier transformation. 
This data are cashed in the memory, and are used as an in-phase signal for the next measurement.

In the software, the reference signal and output signal returned from the NMR coil are multiplied in a "mixer".
The mixer generates two outputs $V_{X}$ and $V_{Y}$;
\begin{eqnarray}
V_{X} &= &A_{s}\sin(\omega t+\theta_{s})\times A_{r}\sin(\omega t+\theta_{r}) \\
        &= &(1/2)A_{s}A_{r}\cos(\theta_{s}-\theta_{r}) \nonumber \\
        &&+(1/2)A_{s}A_{r}\sin(2\omega t+\theta_{s}+\theta_{r})\label{eq_VMIX1}
\end{eqnarray}
\begin{eqnarray}
V_{Y}& = &A_{s}\cos(\omega t+\theta_{s})\times A_{r}\cos(\omega t+\theta_{r}) \\
 &= & (1/2)A_{s}A_{r}\sin(\theta_{s}-\theta_{r}) \nonumber \\
      &&  + (1/2)A_{s}A_{r}\sin(2\omega t+\theta_{s}+\theta_{r}-\pi/2)\label{eq_VMIX2}.
\end{eqnarray}
Since the two inputs to the mixer have exactly the same frequency, the first terms of the mixer outputs in Eqs. (\ref{eq_VMIX1}) and (\ref{eq_VMIX2}) are constant. The second terms have a frequency $2\omega t$. Since the  frequency of the term is doubled, they can be removed using "Integrator". 
Thus, the filtered outputs, $V_{X}'$ and $V_{Y}'$ become
\begin{eqnarray}
V_{X}' \approx (1/2)A_{s}A_{r}\cos(\theta_{s}-\theta_{r}) \\
V_{Y}' \approx (1/2)A_{s}A_{r}\sin(\theta_{s}-\theta_{r}) .
\end{eqnarray}
The noise included in the output signal is removed by the integrator as well.
The in-phase component $I$, the quadrature component $Q$, the signal amplitude $R$,  and the phase $\theta$ are derived from $V_{X}'$ and $V_{Y}'$ by using "Digital Signal Processor" (DSP).
The phase $\theta$ is adjusted by adding a free parameter $\theta_{adj}$ to synchronize with the in-phase signal by using  "Phase Shifter".
All simple formula used by the DSP in the software can be given as,
\begin{eqnarray}
R &= &(2/A_{r}) \times \sqrt{ {V_{X}'}^2 + {V_{Y}'}^2 } = A_{s}\\
\theta & = &\theta_{s}-\theta_{r}+\theta_{adj}=  \tan^{-1}({V_{Y}'}^2 / {V_{X}'}^2)+\theta_{adj}\\
I&= & R \sin (\theta)\\
Q&= & R \cos (\theta).
\end{eqnarray}
It is noted that in these processes mentioned above, we can separate the in-phase component $I$ from the quadrature component $Q$ of the NMR signal.
After repeating many measurements by sweeping the magnetic field, we produce the NMR spectrum.

\section{Evaluation of the portable NMR polarimeter}
\subsection{Experimental procedure}
Since the NMR measurement is carried out without reducing the polarization for the HD target, 
a very weak input signal must be applied.  
Accordingly, an output signal becomes small, which makes the NMR signal observation difficult in noisy circuit background. 
In the NMR measurement, one million samples of digitized NMR signal heights are accumulated and averaged in each magnetic field setting. 
In this averaging process, statistical errors are decreased.
It should be noted that the major part of noises come from thermal fluctuation.
In order to obtain a high $S/N$ ratio, it is important to accumulate a number of  samples within a measurement period at a well controlled constant temperature.
The HD gas with an amount of 1 mol was liquefied in the target cell and was solidified around 17 K in the SC.
The solid HD  was finally cooled down to 1.5 K.
Two kinds of NMR signals were measured with the conventional and portable NMR polarimeter systems by exchanging only polarimeter part (shown in the dashed box of Fig.~\ref{fig:NMR_PXI_diagram.eps}).
The sweeping-speed of the magnetic field was 0.00048 Tesla/s and the RF frequency was kept at 29.45 MHz. 
The magnetic field was swept once and the data were recorded to a hard-disk.

\subsection{Experimental results}
\begin{figure}[hbt]
  \begin{center}
    \includegraphics[width=85mm]{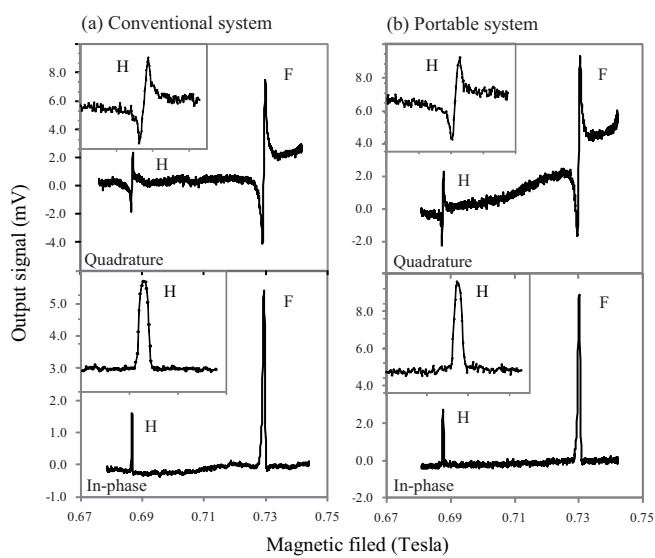}
  \end{center}
  \caption{The NMR signals measured by the (a) conventional and (b) portable systems. The vertical axes are the magnitude of NMR output resonance. The horizontal axes are the magnetic field.}
  \label{fig:NMRspectrum.eps}
\end{figure}

Fig.~\ref{fig:NMRspectrum.eps} shows NMR spectra obtained by using the conventional NMR system and the portable NMR system. 
Both H and F resonance signals are clearly observed at 0.688 and 0.729 Tesla, respectively.
The resonance of F originates from the Teflon (PTFE:Poly-Tetra-Fluoro-Ethylene) coat of silver wire and the target cell and the support frame made of Kel-F$\textregistered$, and that of H originates from the HD.
The F resonance is stronger than the H resonance. 
This is explained by the difference of the numbers of H and F nuclei and the difference of effects of the magnetic field from the NMR coil.
The NMR signals measured with the portable system are stronger than those measured with the conventional system.
This difference would stem from the gain difference of the two systems depending on the RF frequency. 

Table~\ref{Tab:S/N ratios} shows the $S/N$ ratios obtained for H and F  in-phase peaks.
The $S$ is the peak height and the $N$ is the standard deviation of the baseline fluctuation near the peak.
The $S/N$ ratios measured with the portable system are about 50\% lower than those with the conventional system.
\begin{table}[h]
 \caption{$S/N$ ratios of  the H and F NMR signals measured by the conventional and portable NMR systems. }
 \label{Tab:S/N ratios}
 \begin{center}
  \begin{tabular}{|c|c|c|}
    \hline
   &H (proton)&F (Flourine) \\
    \hline
    Conventional system&74$\pm$ 1& 225$\pm$ 1 \\
        \hline
    Portable system& 38$\pm$ 1&111$\pm$ 1 \\
      \hline
        \end{tabular}
 \end{center}
 \end{table}
We checked the stability of the H peak measured with the portable system periodically in a day.
The fluctuation of the peak strength was found to be smaller than 5\%.
This stability satisfies the requirement that the hydrogen polarization degree is determined within a precision of 10\%.

\section{Summary}
A portable NMR polarimeter system was developed for measuring the polarization degree of the polarized HD target at RCNP and SPring-8.
Downsizing of  the NMR system was successfully achieved by building the signal generator, oscilloscope, lock-in amplifier and network analyzer in the software (LabVIEW) with PXI.
The performance of the portable NMR system was proved to be compatible with the conventional NMR system.

The $S/N$ ratio of the portable NMR polarimeter system depends on the mathematical operation speed.
If the number of samplings for averaging increases, statistical errors  decrease.
In the present development, we used a usual laptop PC (CPU:Core 2 Duo T7700, 2.4 GHz) for the portable NMR system. 
Thus, it is expected that the $S/N$ ratio will be improved furthermore if we use PXI-express devices with higher speed of data transportation and a higher performance PC.

\section{Acknowledgments}
The present work was supported in part by the Ministry of 
Education, Science, Sports and Culture of Japan and by 
the National Science Council of Republic of China (Taiwan).
This work was also supported by Program for Enhancing Systematic Education in Graduate Schools and Osaka University.
We thank J.-P.~Didelez, S.~Bouchigny, and G.~Rouille for giving important advice to us. 






\end{document}